# A Reliable Replication Strategy for VoD System using Markov Chain

R. Ashok Kumar
Research Scholar,
School of Computing Sciences
VIT University,
Vellore 632 014, Tamil Nadu, India.

K. GANESAN
Senior Professor
School of Computing Sciences
VIT University,
Vellore 632 014, Tamil Nadu, India.

***Abstract*—In this paper we have investigated on the reliability of streams for a VoD system. The objective of the paper is to maximize the availability of streams for the peers in the VoD system. We have achieved this by using data replication technique in the peers. Hence, we proposed a new data replication technique to optimally store the videos in the peers. The new data replication technique generates more number of replicas than the existing techniques such as random, minimum request and maximize hit. We have also investigated by applying the CTMC model for the reliability of replications during the peer failures. Our result shows that the mean lifetime of replicas are more under various circumstances. We have addressed the practical issues of efficient utilization of overall bandwidth and buffer in the VoD system. We achieved greater success playback probability of videos than the existing techniques.**

## I. INTRODUCTION

The *VoD (Video on Demand)* is one of the most popular services on the Internet. The applications of the *VoD* service are digitally transmitted movies, live streaming video's, distance learning etc., The VOD systems are based on client server architecture. The video server stores the video objects, title, popularity, and the *Quality of Service* (*QoS*) parameters for streaming. In this architecture, a client request for a video to the server and then the server transmits the video streams to the client for playback. The load on the server increases as the number of client request increases. To balance the load on the server, multiple servers are added to the existing VOD System. Popularly these VOD systems are known as parallel [3] and distributed [4] VOD Servers. Each of these servers serves only a group of users rather than all users in the VoD system. The major drawbacks of these systems are the cost of upgrading video server that requires high end servers with additional bandwidth and requires long term working memory capacity. The following approaches are used to reduce the cost factor for upgrading video servers such as a) *Content Delivery Network* (*CDN*) b) *Proxy based system* c) *Peer to Peer system*. In the *CDN* [2] approach videos of the nearer clients are cached and stored in the *Point of Presences(PoP)*. In this approach it reduces the load on the server and the number of network hops but it is not cost effective due to additional hardware and software required at *PoP* to cache the videos. In proxy based approach [5] [6] videos are cached in the proxy server which is cost effective but suffers from scalability problem. In peer to peer approach [7] [8] a video stream relies on application protocol, the peers assist video servers by distributing the streams over the network. In this architecture the bandwidth are shared among peers which reduces the load on the server and the network bandwidth. However, all the peers over the network are not involved in streaming of the videos. This creates a fairness problem among the peers [9].

To further reduce the load on the *VoD* server, a client side caching scheme was proposed known as earthworm [10]. In this scheme the client not only plays the video but also forwards the streams to another client with adequate buffer and delay known as basic chaining. This scheme is further extended as forward, backward, adaptive and optimal chaining which exploits client resources such as buffer and uplink bandwidth [10][11][12][13]. However demand for high quality videos and the longer duration of videos are expected in the nearer future [14]. For such applications the existing chaining schemes fall short in meeting the scalability requirements like bandwidth and buffer.

In our previous work we have defined novel chaining technique known as *M Chaining* [1] which is based on the CTMC model. In this chaining technique, the number of chain increases and reaches the steady state. Once it reaches steady state we found that there was a smooth streaming of movies to the clients.

The above mentioned chaining schemes mainly focus on the server load. To further reduce the load on the server many have investigated on utilization of client resources by combining both Peer to Peer system and chaining mechanism. The Peer to Peer systems have smaller storage capacity and less bandwidth that cannot be a representative of a dedicated central server therefore the redundant movies are stored in the peers to improve the reliability of movies in the peers. Redundancy of movies can be implemented in two different methods one is replication technique and another is Erasure correction code. In replication technique multiple copies of movies are stored in different peers and in Erasure correction code technique a movie is divided into different smaller segments and multiple copies of these segments are stored in different peers.

In our work, we use an architecture which combines Proxy based system and Peer to Peer system to reduce the load on the server. We have used the M-Chaining technique [1] for the transmission of movies in the peers. The problem in this architecture is that peers are unreliable which may frequently break the chain during the transmission of movies between the peers. Hence we propose a new data replication technique on peers for improving the reliability of the movies.

In this paper data replication is defined as the more number of copies of the same video are stored in different peers. We have proposed a new algorithm for data replication to optimally generate the number of replicas. In this paper we have followed four major steps for the overall optimality of the video streams. Firstly the video is replicated into number of replicas and these replicas must be stored in different peers. Secondly it requires a data placement policy, which decides, how many number of replicas should be placed on the peers. Thirdly a selection policy is required to select a peer that contains a replica of the movie for optimal streaming and finally, we have used a *CTMC (Continuous Time Markov Chain*) model to analyze the system performance of the replication in case of peer failures.

The rest of the paper is organized as follows: Section 2 reviews previous related works; Section 3 presents an overview of the proposed *VoD* architecture; Section 4 evaluates the success playback probability and reliability using simulation; and Section 5 concludes the paper.

## II. RELATED WORK

In [15] has proposed a replication technique to maximize the availability of the video, based on the prior information about the availability of movie in the peer and the popularity of the movie. The replicas are generated by maximizing hit rate known as *Max Hit.* They formulate an optimization problem using dynamic programming to maximize the availability of the movie in the peers. The downside of this approach is that the bandwidth of the peer containing the replication of the movie is completely ignored. The movie may be available in serving peer but the requesting peers are completely blocked because of non availability of the bandwidth in the serving peers. In our replication technique, we considered the popularity of the movie and as well as serving peer bandwidth to calculate the number of replicas so that the requesting peers are not completely blocked.

In [15] has proposed another replication technique to minimize the request of the movie based on the currently available bandwidth. The replicas are generated by minimizing the request rate known as *Min Req.* They have formulated this problem using dynamic programming to minimize the request of the movie in the loaded peer. The drawback of this technique is that the replication is done for the most popular movies. As the popularity of the movie reduces, the replication is also reduced. In our replication technique, popularity of the movie is one of the important parameter for increasing the replicas for the most popular movies and to reduce the replicas for the least popular movies. We have streamlined the generation of replicas in such way that even the least popular movie have significant replicas and not worst as *Min Req*.

In [16] has formulated a replication technique based on the availability of the file and proposed a bi-weighted model to find the optimal resource allocation scheme among the files. In this technique, the preference is given to the files that have maximum weightage. But the drawback of this technique, to calculate the weight of the file the author rely on partial and limited information of the file which is located in the neighboring unreliable peers. In our approach, for each movie we have calculated the weight based on the popularity, request arrival and number of replication. The highest weighted movies are given more preference during the placement of the movies in the serving peers. In our case the weight is calculated periodically and modified accordingly in the proxy server so that the non serving does not always depend on the neighboring unreliable peers for the movies. By this way the updated information of the available movie in the neighboring serving peers is obtained frequently from the proxy server so that the non serving peers can switch among the serving peers for the movies in case of serving peer failures.

In [17] has proposed a replication strategy for the super peer that has a request rate table which contains a total number of request and request rate resources for a particular category. The number of replicas is calculated by multiplying the request rate with a $K$ factor, where $K$ is an aggressive replication strategy. The replicas are generated proportional to the request rate and the file size; these replicas are stored uniformly in all heterogeneous peers. The major issue of this strategy is in finding the $K$ factor which is not addressed properly. In our approach, we focus on request arrival rate, down time of the serving peers and currently available bandwidth and buffer of the serving peer to replicate the movies. We also have more number of serving peers because user relinquishes its peer resource information to qualify as a serving peer that also satisfies the minimum resource information. Replicas are generated based on the available resource and request arrival rate in that instance of time instead of multiplying the request rate with some $K$ factor.

In [18] have assumed that all the movies have the same popularity based on *K center problem*. Since this problem is well known *NP complete problem*. When a new request arrives to place a new movie to the super peer, then the new movie is always placed in the super peer. If the storage capacity of the super peer is less than the size of the new movie then some existing movies in the super peer are forcefully removed. In our approach, existing movies are removed based on the *LFU (Least Frequently Used)* algorithm i.e., if the popularity of the existing movie is less than the popularity of the new movie then the least popular movie is removed from the super peer and if the new movie popularity is less than all the existing movies in the super peer then the request to place the new movie is rejected. We have given more preference on popular movies than the non popular movies, so that the popular movies are always available in the super peers.

Durability of the replicas in peers is studied in [19]. The authors concluded that the durability of peers should have a longer life time because of longer repair time. They have defined the lower bound threshold for each super peer as the function of the system capacity and bandwidth. Usually, when the system fails in any circumstances they consider longer duration for the repair time. But this is not the case in our proposal, we have achieved the shorter repair time for stop fail, *VCR* functionality etc., other than hardware and software failures, we have also achieved the greater durability of the replicas in case of some peer failure.

In [20] developed a simple Markov Model to evaluate the parameters of replication system. The parameters for the model is based on replica loss and replica repair. The analysis of the model is coarse because the replication has aggressively maintained in small numbers of replicas in the super peers. Reliability of the replication fails due to the smaller number of replicas and recovery from the failure of the replicas was also not addressed in the problem. Hence in our approach of replication, we have recovered from the failure of the replicas by using Markov Chain.

In [21] discussed the model of birth and death process using *CTMC* for the number of replicas which assumes the independent and exponential failure and repair of the peers. Here the availability and durability are evaluated separately for finding the more accurate model to predict the durability of the system. Insight of this approach seems to produce inaccurate results. In our approach, we have maintained a stronger correlation among the peers to achieve the durability, availability and greater performance of the system.

In [22] is estimated the data durability using *CTMC* model and provided an empirical expression to yield a good approximation of the sub linear parameter values. However the results of the model show some inaccurate predications of the probability of data loss. In our approach, we have focused on finding the accurate prediction of the probability of data loss during the replication of the videos.

## III. SYSTEM MODEL

We have proposed a model that as the combination of proxy based architecture and peer to peer system as depicted in Fig.1. The model contains a main multimedia server, number of proxy server and peer to peer systems. The main multimedia server contains movie files with the following information such as index, size, duration, popularity, minimum buffer, and maximum bandwidth of the movies. The overall system load is equally divided among the clusters. A cluster contains group of peers which are connected to a proxy server. The proxy server contains the streamed movies and currently streaming movies. It also maintains a database of all the peers which are currently available in the cluster. In the peer to peer system a peer can be a serving peer or it can be a non serving peer. A serving peer stores the number of movies and these movies will be served to other non serving peers using a chaining mechanism.

The first problem we have addressed in our system is the identification of a serving peer. Identification is difficult because a peer does not have powerful processing capability when compared to a server. This problem can be solved by the user voluntarily relinquish the peer resource information such as the storage capacity, *CPU* speed and network bandwidth to the proxy server. The peer resource information is stored in the database which is maintained by the proxy server and this peer is designated has a serving peer. Additional details of up time and down time of the serving peers are frequently obtained by the proxy server to maintain the system reliability. In this approach, we find two major benefits one is the overall main multimedia server load is reduced significantly by including these additional bandwidth and buffer of the serving peers in the cluster as well as the user of the serving peer is also benefited by reduction of the subscription fees.

The second problem we have addressed in our system is the placement of movies in the serving peers. Placing of the movie in the serving peer is difficult because of high unreliability and unpredictability of the peer. This difficulty can be eliminated by having more than one duplicated copy of the movies stored in different serving peers with a general rule that no serving peers will have multiple copies of same movie. In case, if a serving peer fails then the copy of the movies of the failed serving peer can also be obtained from other serving peers. In traditional *VoD* system the duplicate copy of the movie is obtained either using *replication strategy* or *erasure coding technique.* In replication strategy the number of replica for the movie is calculated and these replicas are stored in different serving peers. In *Reed-Solomon Erasure Correction Code technique* for *(n-h)* movies we need to calculate *h* redundant movies out of $n$ movies and these $h$ redundant movies are stored in different serving peers. We have observed that erasure code technique generates more copies of duplicate movies and consumes large storage space of the serving peers when compared to replication strategy. Hence, we propose a new replication strategy in our model and the replicas generated from this strategy are stored in different serving peers using *Smallest Load First* algorithm.

The third problem we have addressed in our system is selection of the serving peers. Because of the dynamicity of the serving peers, it is difficult to select a predefined serving peer to serve a non serving peer. In our approach we have used *Least Load First* algorithm to select a serving peer.

The fourth problem we have addressed in our system is to measure the availability of the replicas in the serving peers. The behavior of the serving peers is unpredictable because of the dynamicity of the up time and down time of the serving peers during the streaming session. So we apply *Continuous Time Markov Chain (CTMC)* model to measure the performance of our new replication strategy. Hence we elaborate the detailed discussion of the problem addressed in our model in the next section.

### A. *Initialization of the system and identification of serving peers.*

We consider $N$ is the total number of peers and $G$ is the total number of serving peers in the cluster. $(N\text{-}G)$ is the total number of non serving peers in the cluster. $M$ is the set of movies $\{m_1, m_2, m_3,.. m_m\}$ served in the cluster and $D_m$ is the duration of $m^{th}$ movie and $S_m$ is the size of the $m^{th}$ movie. Each movie $m$ is equally divided in to $V$ number of video blocks such that $m=\sum_{i=1}^{S_m} V_i$. The entire movie follows *Variable Bit Rate* (VBR) for the transmission of video blocks and $C_m$ is the minimum number of channels required for the efficient transmission of the $m^{th}$ movie. $C$ is the sum of all $C_m$ channels required to transmit the movies in the cluster.

The request arrival rate for the $m^{th}$ movie is exponentially distributed with a mean rate $\Lambda_m$ and $q_m$ is the popularity of the $m^{th}$ movie derived from zipf's law. Initially, the proxy server maintains a list of serving peers in its database. This list is created based on the sharing parameters such as requested movie, storage capacity, CPU speed, bandwidth, up time and down time of the serving peer. Whenever a peer requests for a new movie to the main server, the request is redirected to the proxy server in which the peer belongs to that proxy server. The proxy server replies back with a dynamic list of serving peers that contains the requested movie to the requested peer. If the movie is available in the proxy server then the initial portion of the video blocks are directly streamed from the proxy server and the later portion of the movie will be streamed from one of the selected peer from the list of serving peers. If the movie is not available in the proxy server or not available in any of the serving peers then the initial portion of the video blocks are directly streamed from the main server and the later portion of the video blocks are first downloaded and buffered in the proxy server and then streamed from the proxy server to the requested peer. Intuitive behind this idea is all the customers and services providers are profited by optimal utilization of the system resources. Firstly, load on the main multimedia server is reduced because of the load sharing among the peers. We have also found that peers are neither overloaded nor exploited by the chaining mechanism. Secondly, the user of the system is highly benefitted by paying less in their monthly subscription fees. The cost of subscription fees is reduced based on the serving peer's resource utilization and availability. The user's who relinquish their peer as the serving peer is the most beneficiary in this scheme. The users of the non serving peers are also benefited by paying only small percentage of subscription fees than the actual subscription fees. Finally, the overall system resources are utilized very well because it uses only the residual bandwidth and buffer of the serving peers to serve non serving peers.

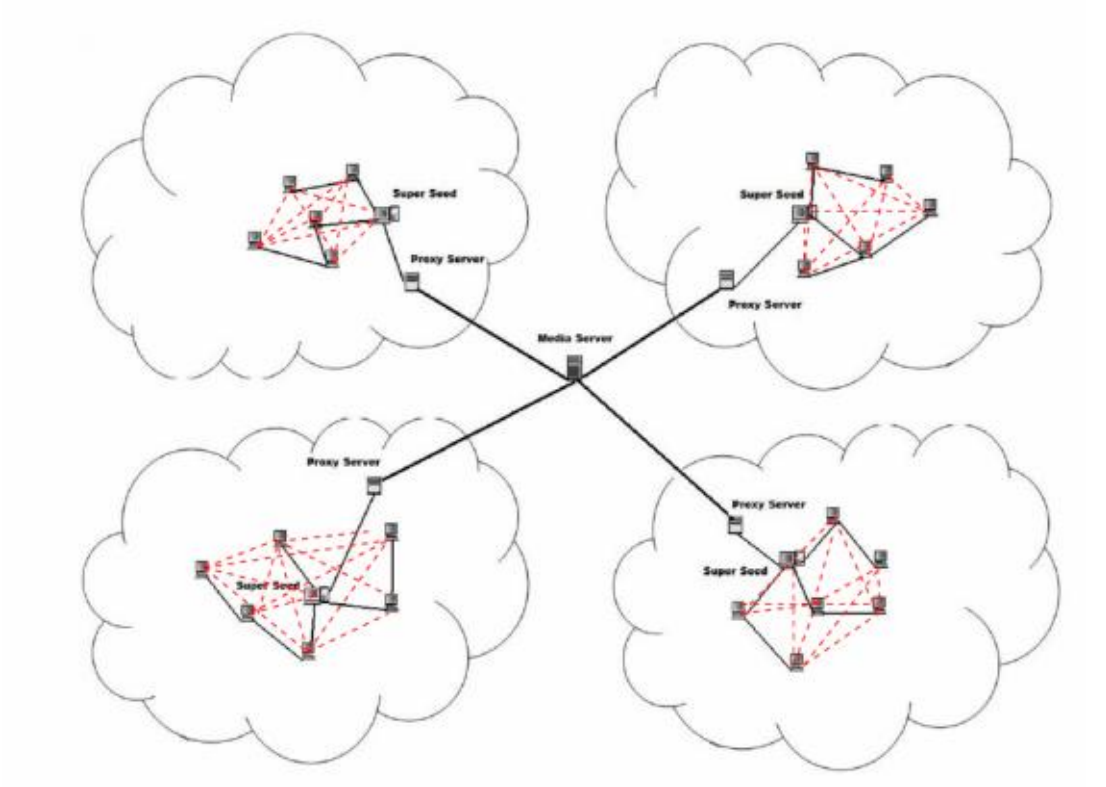

Fig. 1. Proposed Model of VoD System

### B. *Replication Strategy*

Once the serving peers are identified we need to calculate the number of replicas for the movies and place these replicas in different serving peers. The objective of replication strategy is to maximize the availability of the movies in the cluster, in case of serving peer failures. This is achieved by replicating the copies of the movies in the serving peers. A copy of the movie is always stored in the proxy server and duplicated copies are also stored in different serving peers while transmitting the movie in the cluster, which maximizes the availability of the movies in the cluster. We have identified the serving peers as mentioned in section III A, that as sufficient bandwidth and buffer to place additional movies in it. Before placing the movies in the serving peers, the requested movies are downloaded from the main multimedia server and stored in the proxy server. Now we define the different parameters that are required for the replication strategy. Let G be the total number serving peers and G contains the set of currently available serving peers in the cluster. Let $P_{up}$ and $P_{dn}$ is the duration of peer *uptime* and *downtime* respectively of the serving peer in the cluster. The probability of peer availability in the cluster is defined as $A_{up} = \frac{P_{up}}{P_{up}+P_{dn}}$ and the probability of the peer unavailability in the cluster is defined as $A_{dn} = \frac{P_{dn}}{P_{up}+P_{dn}}$. A flag is associated with each of the serving peer based on the value obtained from $A_{up}$ and $A_{dn}$ . If the value of the flag is 1 then it indicates that the serving peer is available and it can store the replica of the movie. Otherwise the serving peer is not available and it cannot store the replica of the movie. The reason for unavailability of the serving peer can be software or hardware fail, stop fail, network failure, insufficient network bandwidth or not enough storage space to store replicas. We have assumed that peer unavailability is much larger than peer availability in our model.

Each serving peer $g$ shares $B_g$ bytes of storage space and $\beta_g$ channels of uplink bandwidth. Let $R_m$ be the number of replicas for the $m^{th}$ movie and $R$ be the total number of replicas in the cluster. A valid replication should not exceed the aggregate storage capacity of all the serving peers available in the cluster such that $\sum_{m=1}^{M} S_m R_m \le \sum_{g=1}^{G} B_g$ and should also satisfy the channel requirement such that $\sum_{g=1}^{G} \beta_g \ge C$ for the transmission of the movies. Now we propose a new replication algorithm for the proxy servers to generate optimal number of replicas for a cluster.

*Replication Algorithm in Proxy Server*

Step 1: Wait in batch for the new request arrival of the movies.
Step 2: Sort the movies in the batch based on the popularity $q$.
Step 3: Measure the current request rate $\Lambda$ for each of the new requested movies.
Step 4: Get the size of each requested new movie file from the main multimedia server and store in a new list $S_m'$.
Step 5: Get the list of serving peers with $A_{up}$ flag as 1 from G and store in a new list G′.
Step 6: Filter out the serving peers from G′ based on $B_g < S_m'$ and store the remaining serving peers in a new list G″.
Step 7: Total number of available serving peers $T_R$ = count(G″).
Step 8: $\Omega = \text{avg}(\Lambda, q)$
Step 9: Replicas $R = \Omega * T_R$
Step 10: Choose $R$ number of serving peers from G″ and store in a new list G‴.
Step 11: Placement ($R$, G‴ )
Step 12: Repeat from Step 1.

In our *proposed* replication algorithm, we had considered the other important factors such as peer uptime for reliability, popularity of the movie and inter arrival request rate of the movie. The replication algorithm increases the number of replicas for the most popular movies and balances the load across all the serving peers.

### C. Placement of replicas in the serving peers

After determining the $R$ number of replicas we must place these replicas in the serving peers without overloading them. We can select either a *Random* technique or a *Round Robin* technique to place the replicas in the serving peers. These placement techniques assume all movies to have same popularity that is not the case in the current scenario of VoD systems. We have to emphasis more on the popularity of the movie because of the demand created in the system. Hence, we have used *Smallest Load First* algorithm to place the replicas of the movies in the serving peers. In this algorithm we have calculated the weight of each movie so that the highest popular movies will have larger weights when compared to its counter part of the movies with the least popular.

*Smallest Load First* algorithm

Step 1: Accept number of replicas $R$ and list of active serving peers G‴ from the *Replication Algorithm.*
Step 2: For each of the movie in $R$ calculate the weight based on the request arrival rate, popularity and the number of replicas. The weight of each movie is calculated by $W_m = \frac{\Lambda_m X q_m}{R_m}$. Map each movie with associated weight and store in $\dot{M}\dot{W}$ hash table.
Step 3: Sort all movies stored in $\dot{M}\dot{W}$ based on their weights in non ascending order.
Step 4: For each iteration $Q$ ($Q$ is the number of replicas that a serving peer can store in it.)
Select $Q$ number of replicas from $\dot{M}\dot{W}$ with the highest weight and place in $Q$ different serving peers subjected to that the highest weighted movies are placed in smallest load first serving peers and no multiple copies of the same movie are stored in the same peer.

### D. Selection of a Serving Peer

After placing the replicas of movies in different serving peers, next step is to select a serving peer to serve a non serving peer. Whenever a non serving peer makes a request for a movie with its existing resource information to the proxy server then proxy server replies to this non serving peer with the list of all active serving peers which contains the requested movie in it. Now the non serving peer should make a decision in selecting a serving peer from the list. One method of choosing a serving peer is to select a serving peer randomly. The drawback of this method fosters the load imbalance among the serving peers. Hence the non serving peer must check for the least loaded serving peer to receive the requested movie. Now the non serving peer makes a request of currently available resource information to all the serving peers in the list. Then the requested non serving peer receives currently available resource information from all the requested serving peers and it sorts all the requested serving peers in non-increasing fashion based on the available resource of the requested serving peers. Finally the non serving peer will select a serving peer with the highest resource availability and least load for the reception of the movie. During the transmission session the serving peers can fail at any time. If the serving peer fails then the non serving peer again executes the same procedure to select a serving peer with a highest available resource and least loaded serving peers. In the worst case scenario if none of the serving peers are available then the movies are directly transmitted to the non serving peer from the proxy server.

### E. *Reliability Model.*

In this section we have formulated a reliability model for the replicas of the movies. The objective of this section is to maximize the availability of popular movies in the serving peers. As we have already discussed the replication strategy in section 3 B and revealed that more number of replicas is generated for the most popular movies. Now, we have the problem of availability and durability of the replicas in serving peers. This problem raises certain questions regarding the availability and durability of the replicas in the serving peers. How long the serving peers can participate in the process of the replication? What is the reliability of serving peer up time during the replication? What happens if the serving peer fails abruptly? How often the peers join and leave the system as a serving peer? To solve this problem we apply *CTMC* model to normalize the number of replicas during the transmission of movies. We have considered $n$ states of serving peers in the *CTMC* model as shown in Fig. 2, where $k$ is the functioning of replicas for the $m^{th}$ movie. The state 0 is the absorbing state beyond which there is no replication. We relate our problem with the *Gambler ruin* problem, such that the probability of a state contains certain number of replicas during transmission can fail and it can be ruined due to non availability of replicas in the long run of the system. Hence we evaluate the state duration $T_p$, where $T_p$ is assumed to be exponentially distributed with mean $1/\lambda$, where $\lambda$ is the failure rate. Thus the reliability is defined as $P[T_p > t] = e^{-\lambda t}$ at any given time $t$. Over the period of time, the peer fails and decreases the number of replicas in the system. To address these attrition; we must have a repair mechanism to create new copies of replicas during the peer failure. The repair must identify the lost replica and copy the lost replica to the existing serving peer. This process takes some duration $T_r$, where we assume that $T_r$ is also exponentially distributed with mean $1/\mu$, where $\mu$ is the repair rate. Thus the reliability is defined as $P[T_r > t] = e^{-\lambda t}$ at any given time $t$. To balance between the fast failure rate and fast creation of new replicas, we define a normalize function of *repair rate* $\gamma$ as $\mu / \lambda$. The repair time $T_r$ must be at least the time it takes to detect the lost replicas and copy of the new replicas to another serving peer.

### F. *Markov chain*

The above model is analyzed and reduced to Markov chain. The system has $k$ functioning replicas at any given point of time and the remaining $(n-k)$ replicas are being repaired, this system can be modeled as markov chain having $(n+1)$ states. In state $k$, any one of $k$ functioning replicas can fail, in which case it transits to $(k-1)$ state or any one of $(n-k)$ non functioning replicas is repaired, in which case it transits to $(k+1)$ state. According to the CTMC model the transition is defined as follows. In state $k$ the transition can occur either to the state $(k-1)$ with rate $k\lambda$ or to the state $(k+1)$ with rate $(n-k)\mu$. Note that state 0 is the absorbing state beyond in which there is no replication.

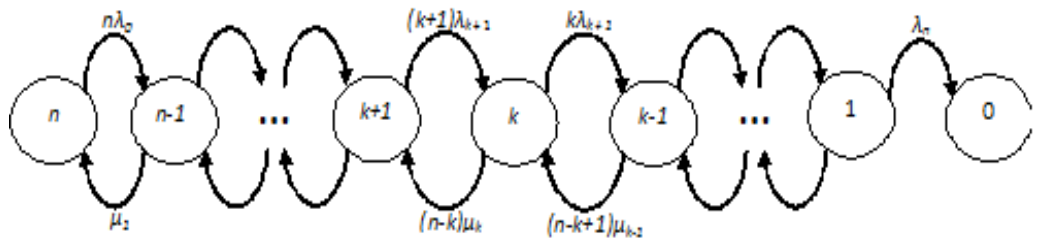

Fig.2. CTMC Model

### G. *Mean time to failure*

Due to the dynamics of peers in the cluster, we need to calculate the life time of replicas before it is lost permanently. The relevant metric in this model is the time to failure i.e the time taken from $n$ replicas of a state to reach the absorbing state 0 with no replicas left. We define the life time of replicas as $T_s$ as the product of expected time $n_e$ and expected time $t_e$ where $n_e$ is the number of state it traverse and $t_e$ is the total duration spend in each state. We define expected life time of replicas in equation $T_s = n_e * t_e$. We derive the values of $n_e$ and $t_e$ as follows. Let $Q_k$ be the probability that the system reaches the absorbing state 0 starting from state $k$ before state $n$. $Q_k$ satisfies the following recurrence relation, where $0 < k < n$

$$Q_k = \frac{k\lambda}{k\lambda + (n-k)/\mu} Q_{k+1} + \frac{(n-k)/\mu}{k\lambda + (n-k)/\mu} Q_{k-1}$$

$$Q_k = p_k Q_{k-1} + q_k Q_{k+1}$$

Let $p_k + q_k = 1$ where

$$p_k = \frac{k\lambda}{k\lambda + (n-k)/\mu} \; and \; q_k = \frac{(n-k)/\mu}{k\lambda + (n-k)/\mu}$$

$$(p_k + q_k) Q_k = p_k Q_{k-1} + q_k Q_{k+1}$$

$$(Q_{k-1} - Q_k) = \frac{q_k}{p_k} (Q_k - Q_{k+1})$$

$$(Q_{k-1} - Q_k) = \frac{n-k}{k} (\mu/\lambda) (Q_k - Q_{k+1})$$

We then start from state $(n-1)$, to calculate the probability of traverse rate $Q^*$ to reach the absorbing state 0 is given by the following recurrence relation

$$Q^* = \sum_{k=0}^{n-1} \frac{1}{\binom{n-1}{k}} \left(\frac{\mu}{\lambda}\right)^k$$

Therefore, the expected value for the number of state traversed is given by $n_e=1/Q^*$.

Let $T_k$ be the expected time before the system reaches the absorbing state 0 staring from state $k$ before state $n$. $T_k$ satisfies the following recurrence relation, where $0 < k < n$

$$T_n = \frac{1}{n\lambda_1 + \mu_1} + T_{n-1}$$

$$T_{n-1} = \frac{1}{(n-1)\lambda_1 + \mu_1} + \frac{(n-1)\lambda_1}{(n-1)\lambda_1 + \mu_1} T_{n-2} + \frac{\mu_1}{(n-1)\lambda_1 + \mu_1} T_n$$

$$T_{n-2} = \frac{1}{(n-2)\lambda_3 + 2\mu_2} + \frac{(n-2)\lambda_3}{(n-1)\lambda_3 + 2\mu_2} T_{n-3} + \frac{2\mu_2}{(n-2)\lambda_3 + 2\mu_2} T_{n-1}$$

•
•
•

$$T_k = \frac{1}{k\lambda_{k+1} + (n-k)\mu} + \frac{k\lambda_{k+1}}{(k\lambda_{k+1}) + (n-k)\mu} T_{k-1} + \frac{(n-k)\mu}{(k\lambda_{k+1}) + (n-k)\mu} T_{k+1}$$

which yields,

$$T_{n-1} = \frac{1}{(n-1)\lambda_1} + T_{n-2} + \frac{\mu_1}{(n\lambda_1 + \mu_1)(n-1)\lambda_1}$$

$$T_{n-2} = \frac{1}{(n-2)\lambda_3} + T_{n-3} + \frac{2\mu_2}{(n-2)\lambda_3} + \frac{2\mu_2\mu_1}{(n-2)\lambda_3(n\lambda_1 + \mu_1)(n-1)\lambda_1}$$

$$T_k = \frac{1}{(n-k+1)\lambda_k} + \frac{k\mu_{k-1}}{(n-k+1)\lambda_k + (n-k)\mu} T_{k-1} + \frac{k\mu_k\mu_{k-1}}{(n-k+1)\lambda_k + (n\lambda_k - \mu_k)(n-k)\lambda_k}$$

We now start from *(n-1)* state to calculate the probability of the expected time $T^*$ to reach the absorbing state 0 is given by the following recurrence relation.

$$T^* = \left[\sum_{j=1}^{n} \frac{\prod_{k=0}^{j-1} k\mu_{k-1}}{\prod_{k=0}^{j} (n-k+1)(n\lambda_k + \mu_k) + (n-k)\lambda_k}\right]$$

Therefore, the total time spent in each state is given by $t_e=1/T^*$.

In order to maximize $T_s$, the system should increase repair ratio or number of replicas as large as possible. But the problem is how to increase the value of repair ratio $\gamma$ or number of replicas with storage capacity $\eta$ relatively large. Therefore, we consider the effect of some factors to choose the value of $\eta$ and $\gamma$ to maximize $T_s$. The following are the factors we considered in our system. a) *Storage capacity:* The number of replicas created for a movie is limited by the total storage capacity of the system such that $\eta <= \eta_{max}$ where $\eta_{max}$ is the upper limit on the number of replicas due to storage limit. b) *Detecting replica loss:* This factor directly effect on the replica repair time $T_r$ such that $T_r <= \tau_{max}$ where $\tau_{max}$ is upper limit on the normalized repair rate. c) *Repair bandwidth:* Final factor we considered is the bandwidth constraint to create new replicas in the system. A replica exists in a peer for a duration $T_p$ and after which peer departs. The repair mechanism then creates a new replica after time $T_r$. The new replica is created on an average time of $E[T_p+T_r]=1/\lambda +1/\mu$. Starting with $\eta$ number of replicas with $B$ bytes, the average bandwidth is defined by $\varphi=\eta/(\gamma+1/\gamma)$, such that $\varphi <= \varphi_{max}$ where $\varphi_{max}$ is the bandwidth constraint in terms of bytes per second.

## VI. SIMULATION AND RESULT

In this section, we use simulation to evaluate the performance of the proposed technique and compared the results with the existing techniques. We used *MATLAB* software to evaluate the performance of the system. The result of the simulation is evaluated with different success playback probability within the cluster. The success playback probability is defined for a video with length *t*, is the probability of successful reception throughout the entire duration of the video given that *VoD* request is admitted. We also evaluated the mean life time of the replicas to observe the availability of movies in the cluster. The topology used in the simulation is a single media server and 5 cluster based network. Each cluster constitutes a proxy server and 1500 peers, which includes both serving and non serving peers. The proxy server consists of bandwidth that ranges from 30MB to 120MB and buffer ranges from 1000MB to 5000GB. The proxy server maintains a database of currently streamed/streaming movie and a list of serving peers within the cluster. The total number of movies requested in a cluster is less than 300 movies per hour. The length of the movie is 7200 secs and follows a variable bit rate for the transmission. The request arrival rate follows the Poisson distribution. Each serving peer has the twice the uplink bandwidth of its counterpart non serving peer and can store upto 10 movies. The popularity of the movie follows the *Zipf's* distribution with skew factor of 0.271. The peer follows an exponential distribution with mean 3600 secs and 32400 secs for uptime and downtime respectively.

The simulated model is evaluated several times. The result shown is an average of all simulation trials is carried out in all clusters. We simulated different replication strategies namely random, minimize request (MinReq) and Maximize hit ratio (MaxHit) and also compared with our proposed technique.

Fig. 3 shows number of replicas for each movie. The lower index numbered movie is the most popular and higher index numbered movie is least popular. In random replication technique the replicas of the movie are randomly distributed. In maximize hit rate replication technique the replicas are almost equally distributed for each movie. We have observed in MinReq technique more number of replicas is created for the most popular videos and decreases linearly as the popularity of the video decreases. Our proposed technique is much better than other techniques. As we observed that the more number of replicas are generated for most popular videos and decreases linearly to the least popular videos without reducing many replicas for the least popular videos. Our technique is much better in replication because it generates more number of replicas for the most popular videos and not much replicas for the least popular videos when compared to other techniques.

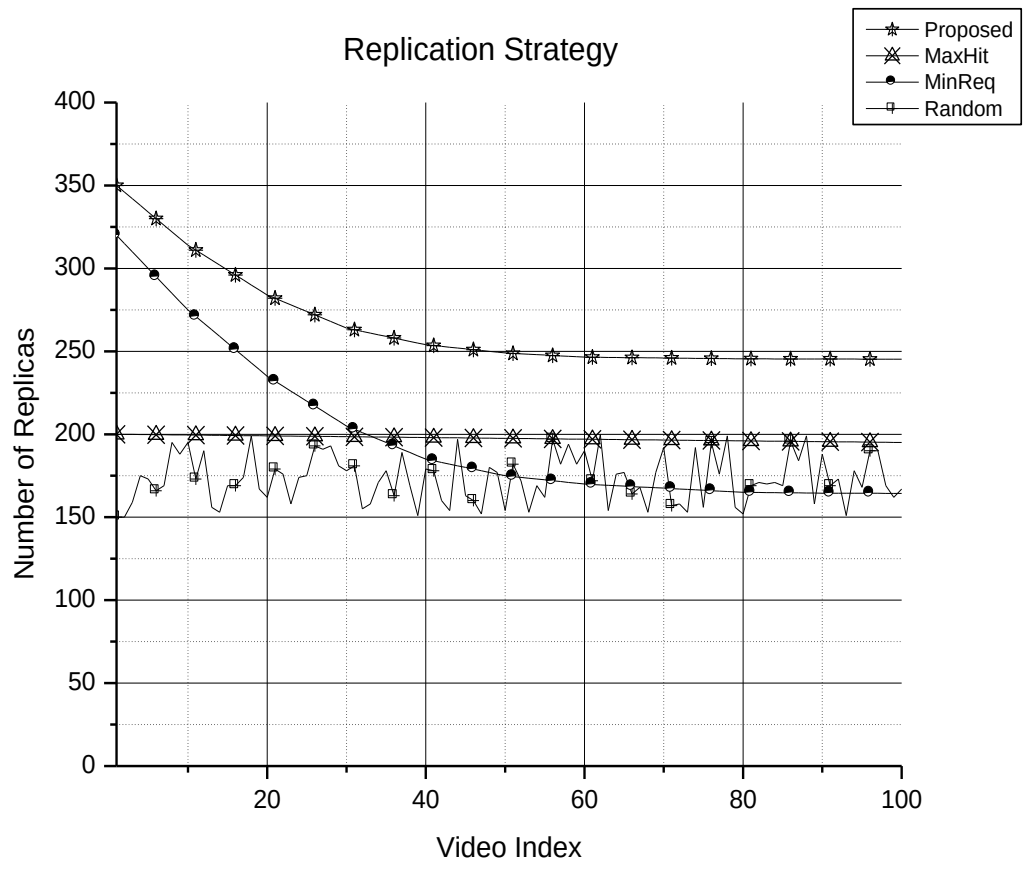

Fig. 3. Comparison of Replication Strategy

Fig. 4 shows the success playback probability of movies with respect to poison arrival rate. Initially, when the arrival rate is less, then we get 100% successful playback. As the arrival rate increases with respect to time the success playback probability decreases linearly. As we observed that our proposed technique achieves success playback probability is 8% more than minimize request and 15% more than maximize hit and 18% more than random technique.

Fig. 5 shows the success playback probability with respect to availability of serving peers. We plot success playback probability for serving peers availability ranging from 0.05 to 0.25. We observed that as the availability of serving peers decreases the success playback probability is also decreases we get 99.8% success playback probability when the availability of the serving peers reaches 0.24 out of all *G* serving peers. We observe our proposed technique achieves the success playback probability of 4% more than minimize request and 16% more than maximize hit rate and 24% more than random technique.

Fig. 6 shows the mean lifetime for the number of replicas available in the system. Using our proposed replication technique, we varied the repair ratio $\gamma$ as discussed in section III E, and calculated the mean life time of replicas. As the value of $\gamma$=0.1 the life time of replicas are less when compared to the value $\gamma$=10.0, on contrary we evaluate the mean of life to different number of replicas. As we observed in Fig. 7, when repair ratio $\gamma$ decreases, the life time of the replicas also decreases and when we increase the repair ratio $\gamma$ the life time of the replicas increases exponentially.

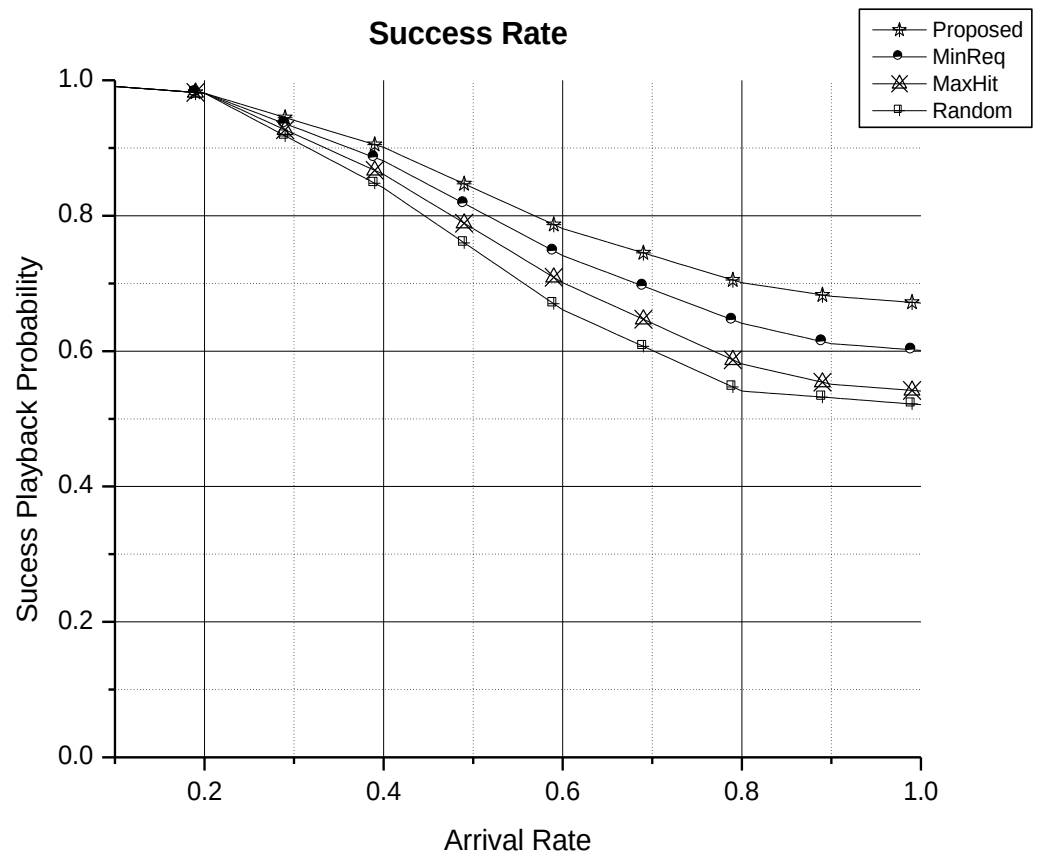

Fig. 4. Success Playback Probability of Movies.

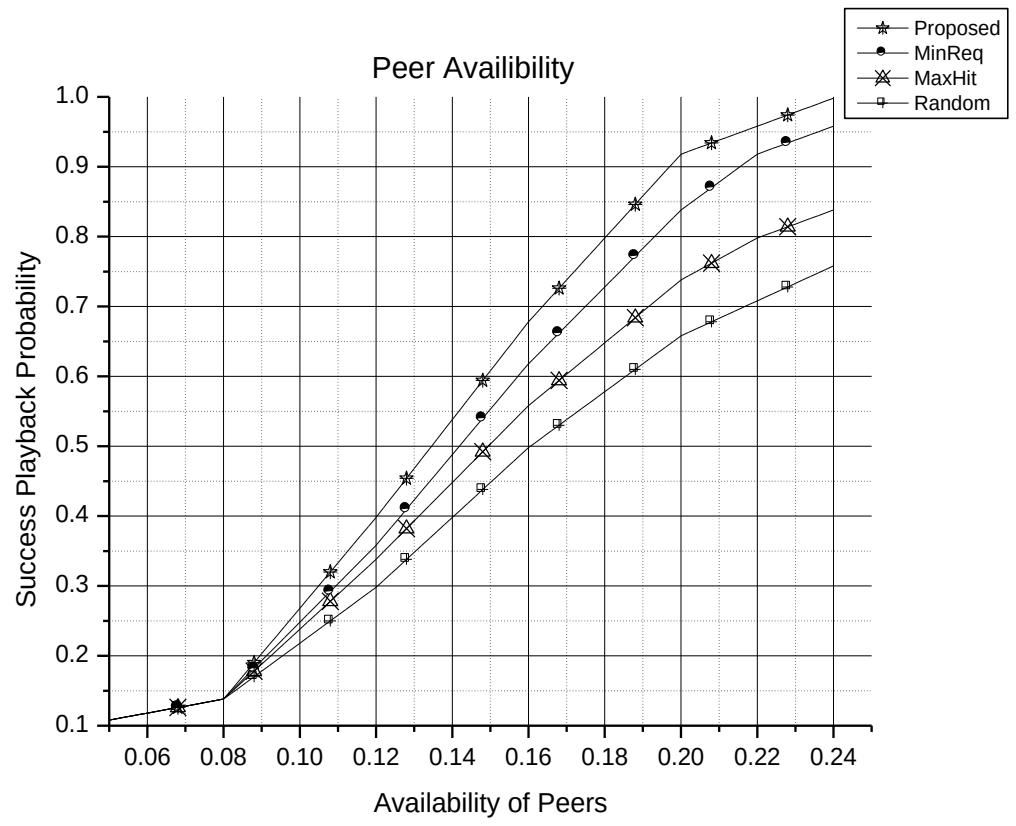

Fig. 5. Availability of Serving Peers.

The success rate of the VoD system is shown in the Fig. 8. We can observe the figure that most of the requested movies are served immediately either from the proxy server or serving peers. If it is served from the serving peer then there is some latency in chaining as shown in the Fig. 8. The rejections of the movies are very less. Fig. 9 and Fig. 10 show average buffer and bandwidth utilization in the proxy servers. As we observe from the figure, initially more number of buffer and bandwidth is utilized. As time proceeds we observed that there was an enormous decrease in the utilization of the proxy server's bandwidth and buffer due to chaining among the peers in the cluster.

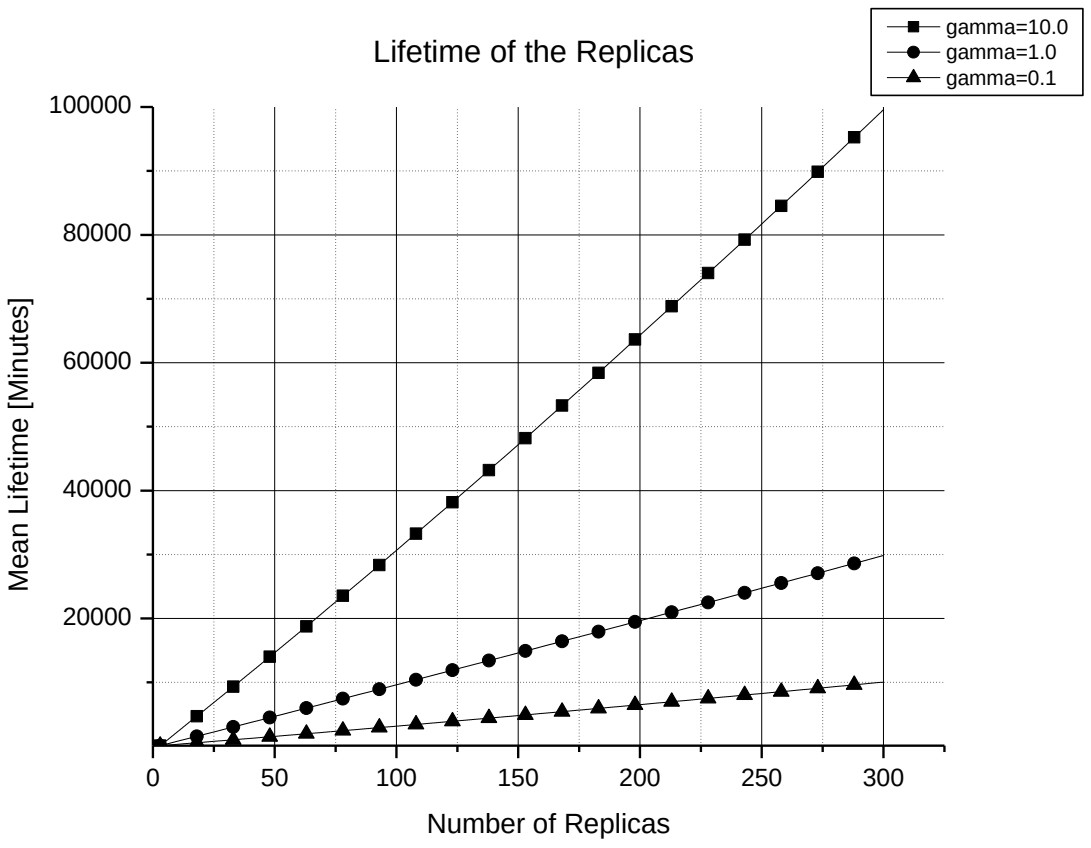

Fig. 6. Mean Lifetime of the replicas

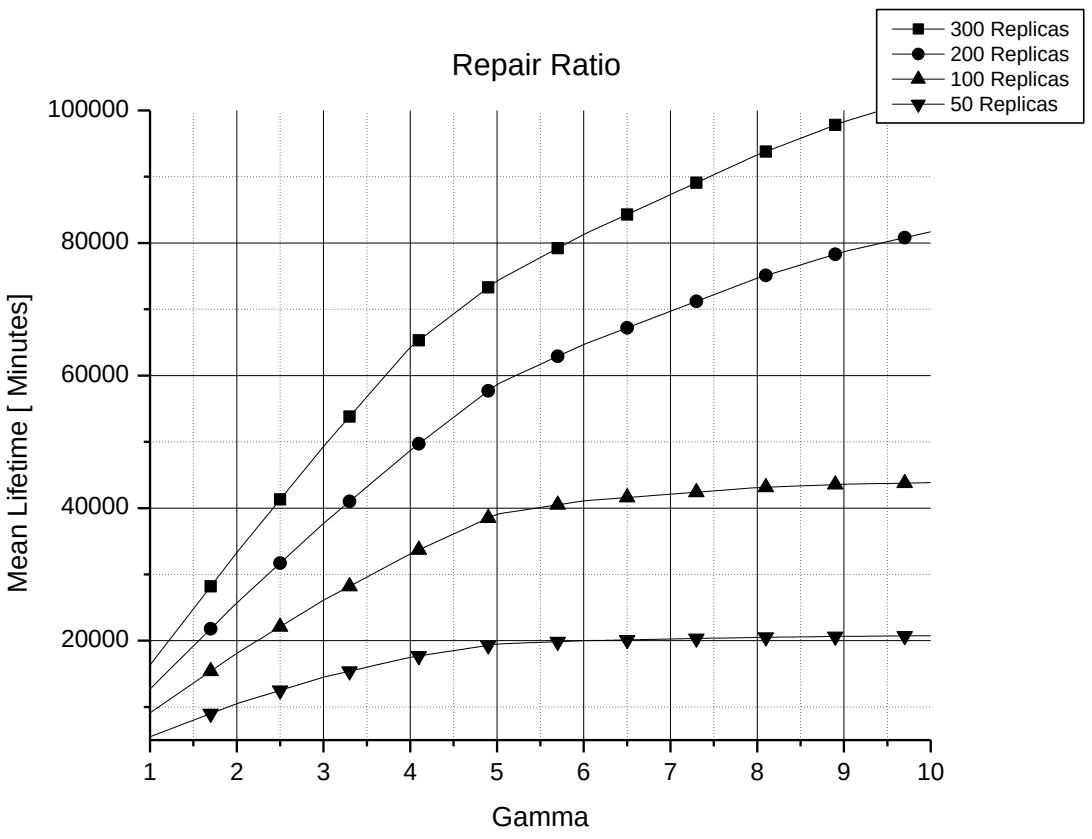

Fig. 7. Mean Lifetime of the replicas based on repair ratio

## V. CONCLUSION

In this paper, we have studied the different types of data replication techniques and proposed a new novel replication technique to generate more number of replicas. We addressed the issues of identifying a peer, generating the replicas, placing the replicas in the peers, selecting a serving peers and reliability of replicas. In particularly, we have investigated intensely for the reliability of replicas in the VoD system. Hence, we have applied the CTMC model and analyzed the reliability of replicas. Our simulation shows promising results in generating more number of replicas and achieving the greater success playback rate compared to the existing techniques. We have also proved, greater lifetime for the availability of replicas in the peers but with the cost of short latency in copying replicas of the failed peer to a new serving peer. Finally, we have achieved the most efficient way of utilizing the overall bandwidth and buffer in the VoD system. Further the paper can be improved by comparing with RSE technique for the better results.

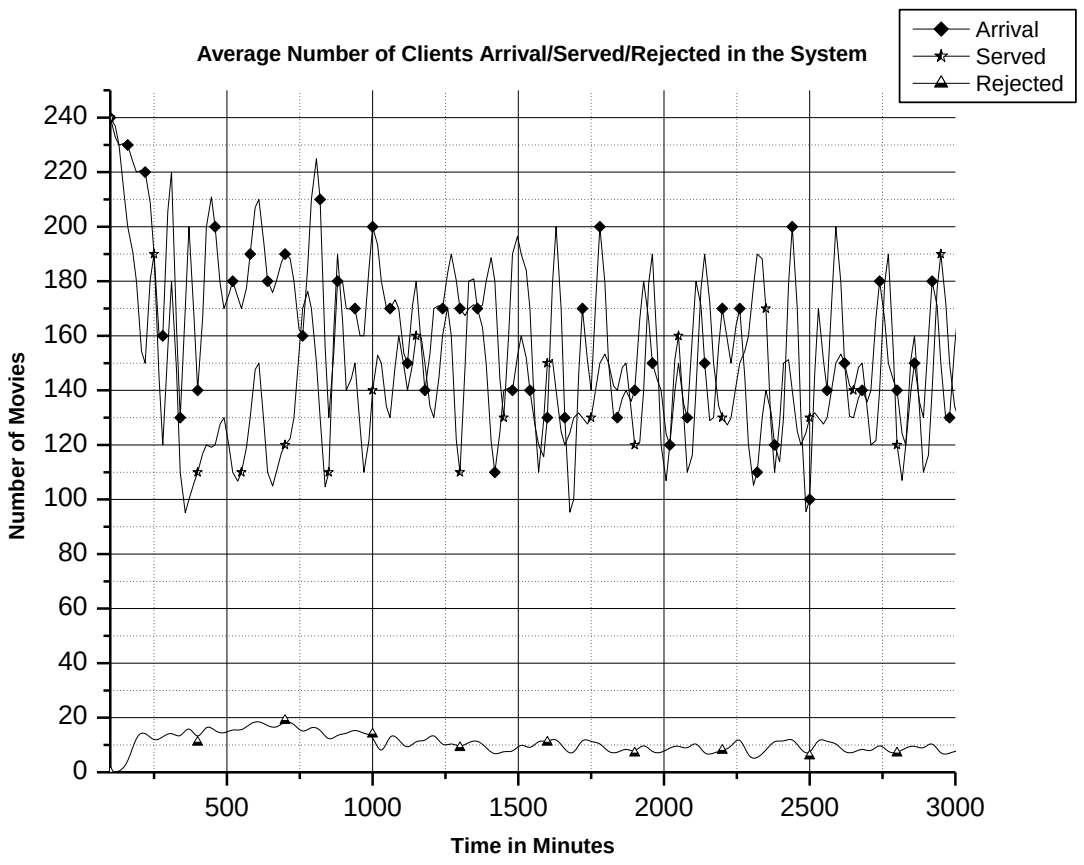

Fig. 8. Success rate of the VoD System

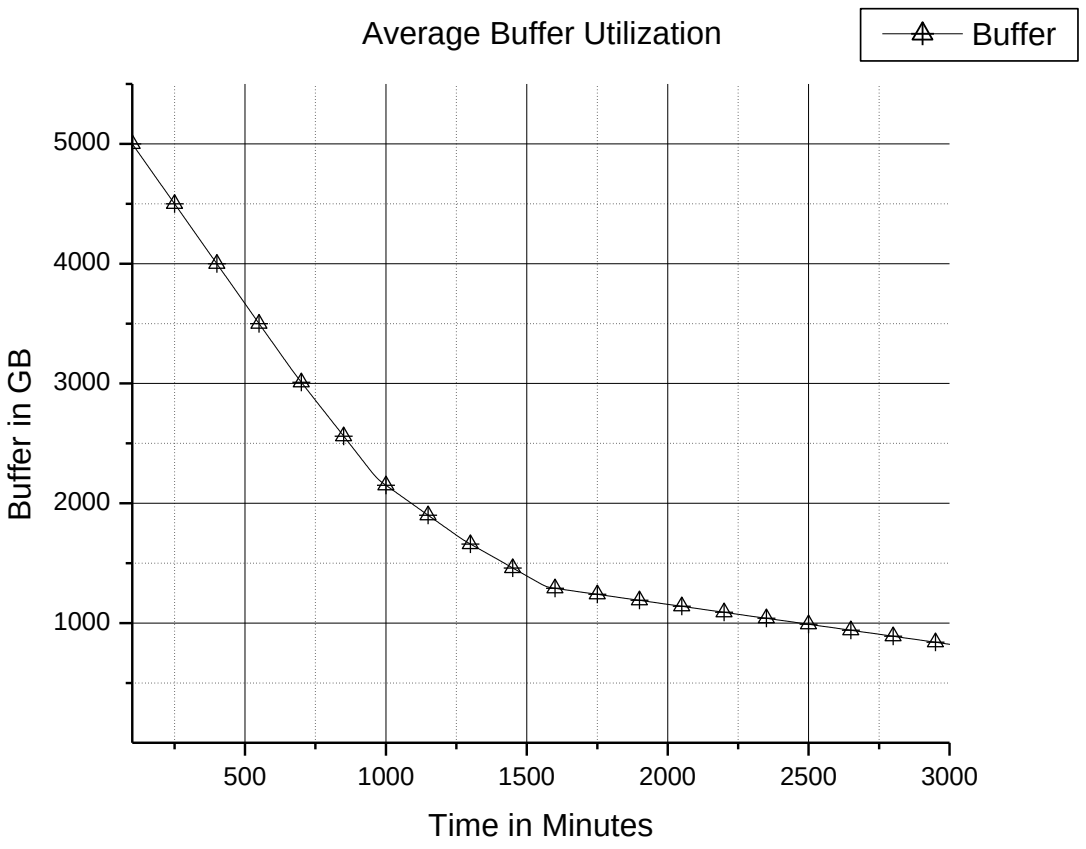

Fig. 9. Average Buffer Utilization in proxy servers.

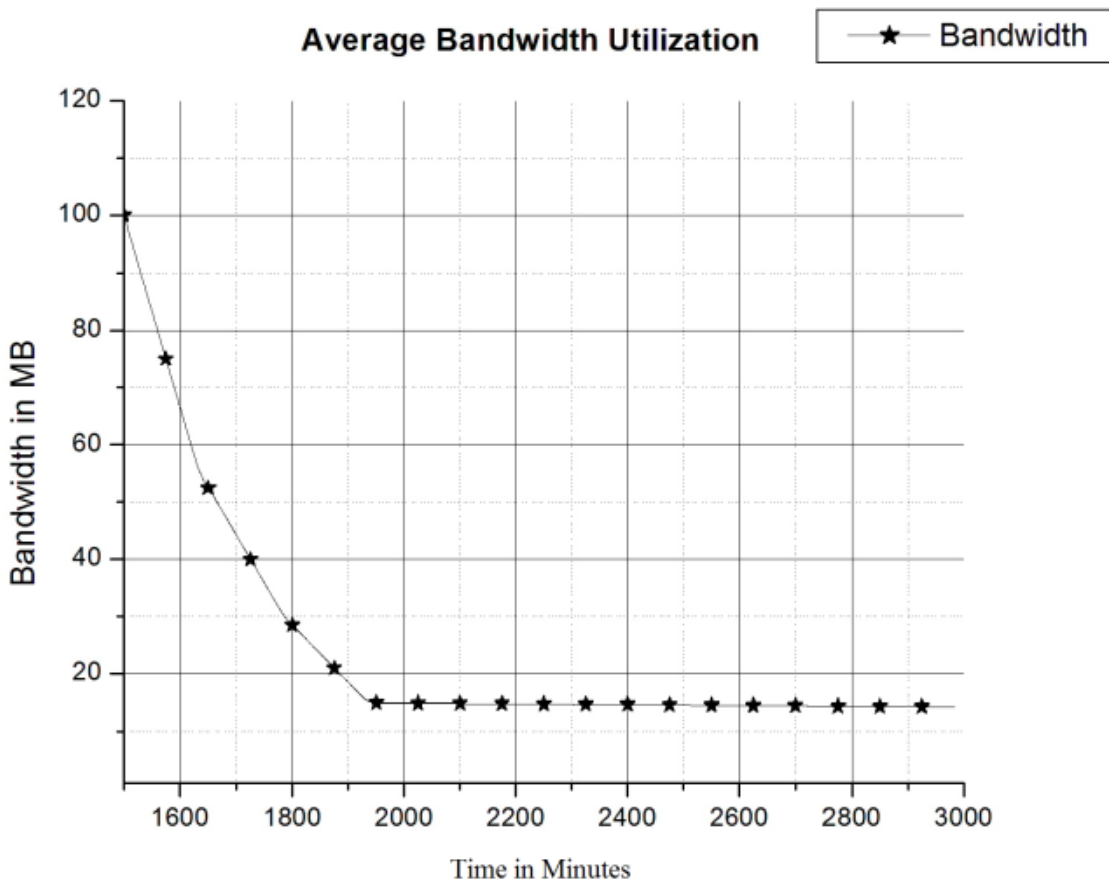

Fig. 10. Average Bandwidth Utilization in proxy servers.

## AUTHORS PROFILE

**R Ashok Kumar** is a Research Scholar in the School of Computing Science at VIT University, Vellore, India. He received his BE degree in Information Science & Engineering from Bangalore University, and MTech in Computer Science & Engineering from Visvesvaraya Technological University, Belgaum, India. His area of interest is Multimedia Applications, Database Management Systems, Internet Technologies and Quality of Service. His current research includes Video on demand systems and Bandwidth Management.

**Dr. K. Ganesan** is currently working as a Senior Professor in the School of Computing Sciences, VIT University, Vellore, India. He is also currently heading the Centre of Relevance and Excellence in "Automotive Infotronics" at VIT University, Vellore. He is also currently working on a Defense Research projected related with "Cryptography". He has published about 50 papers in International journals and conferences (National and International level). He has been honored as the **International Scientist of the Year 2008** by the International Biographical Centre of Cambridge, England. He has been chosen as a candidate for inclusion in the 10$^{th}$ Anniversary Edition of **Marquis Who's Who in Science and Engineering.** His areas of research include Media processing (Image and Video), Mobile Computing and Data security.